\documentclass[%
prb,%
amsmath,amssymb,
reprint,%
]{revtex4-1}
\usepackage[english]{babel}
\usepackage[latin1]{inputenc}
\usepackage{graphicx}
\usepackage{subfigure}
\usepackage{epstopdf}
\usepackage{dcolumn}
\usepackage{bm}
\usepackage{stackrel}
\usepackage[sort&compress]{natbib}
\makeatletter

\begin{document}

\title{Observation of excitonic resonances in the second harmonic spectrum of MoS$_2$ }

\author{Mads L. Trolle, Yao-Chung Tsao, Kjeld Pedersen and Thomas G. Pedersen}

\address{Department of Physics and Nanotechnology, Aalborg University, Skjernvej
4A, DK-9220 Aalborg East, Denmark}
\begin{abstract}
We measure the second-harmonic (SH) response of MoS$_2$ in a broad fundamental photon energy range between 0.85 and 1.7 eV, and present a continuous SH spectrum capturing signatures of both the fundamental A and B excitons in addition to the intense C resonance. 
Moreover, we interpret our results in terms of the first exciton simulation of the SH properties of multi-layered MoS$_2$ samples, represented by a study of trilayer MoS$_2$. 
The good agreement between theory and experiments allows us to establish a connection between the measured spectrum and the underlying electronic structure, in the process elucidating the non-linear excitation of electron-hole pairs.
\end{abstract}
\maketitle

In recent years, two-dimensional transition metal dichalcogenides (TMDs), being semiconductors\cite{Mak2010,Splendiani2011,Eda2011,Zhang2014} with band gaps in the 2-3 eV range\cite{Komsa2012,Qiu2013,Zhang2014,Wang2014}, have been applied as the basis for a host of novel two-dimensional optoelectronic devices\cite{Radisavljevic2011,Yin2012,Larentis2012,Georgiou2013,Chuang2014,Jo2014}.
Accordingly, a great deal of attention has been devoted to the linear optical properties of these compounds\cite{Mak2010,Splendiani2011,Eda2011}. Also, non-linear optical techniques have been demonstrated as particularly powerful probes for the microscopic structure of few-layered TMD crystals, revealing e.g., their crystallographic orientation\cite{Malard2013,Kumar2013,Heinz2013,Janisch2014,Wang2014}, and stacking angle\cite{Hsu2014}. Moreover, second-harmonic (SH) spectroscopy allows for insights into the electronic structure of TMD flakes\cite{Malard2013}, exposing edge-localized effects\cite{Yin2014}, or valley-coherent excitations\cite{Wang2014}.

While experimentally determined linear response functions of various TMDs, dominated by electron-hole pairs bound by several hundred meVs\cite{Komsa2012,Qiu2013,Chernikov2014}, are reproduced reasonably well at the Bethe-Salpeter equation (BSE) level of complexity\cite{Molina2013,Qiu2013,Trolle2014,Gruning2014}, similar comparisons for non-linear cases are lacking. Although several proposed theoretical models, based on both independent-particle\cite{Yin2014,Guo2014} and excitonic\cite{Trolle2014,Gruning2014} approaches, have been published, 
the SH spectrum 
of MoS$_2$ has been investigated experimentally\cite{Malard2013} 
 only in a narrow fundamental photon energy range between 1.2 and 1.7 eV, probing the so-called C resonance also known from linear optics. Theoretical models have so far been benchmarked by their ability to reproduce this  single feature\cite{Trolle2014,Gruning2014,Guo2014}.
 However, the parabolic dispersion, and the split valence bands near the K-points of the Brillouin zone, translates into bound electron-hole pairs in the BSE picture. These give rise to sharp peaks in the absorption spectrum due to the so-called A and B excitons\cite{Qiu2013,Li2014} at photon energies near 1.9 eV, and the question remains how they might affect the SH response. In particular, the relative intensities and lineshapes of such features are of interest.

In the present letter, we report an experimental optical SH spectrum generated from many-layered MoS$_2$, with fundamental photon energies varied in a broad range between 0.85 and 1.7 eV, capturing both the fundamental A and B excitons in addition to the aforementioned C resonance.
%
%
We perform our optical experiments on (i) many-layered MoS$_2$ flakes exfoliated onto fused silica by the well-known scotch tape method\cite{Radisavljevic2011}, and (ii) a bulk MoS$_2$ crystal. 
The first approach gives access to a relatively smooth surface suitable for optical experiments, with little diffuse scattering. However, here flake thickness and surface coverage are difficult parameters to control within the area of a mm-sized laser beam. The thickness of the bulk crystal, on the other hand, is immaterial due to absorption. Unfortunately, the rather rough surface of weakly van der Waals bound bulk TMD crystals typically make systematic optical studies difficult. 
Hence, due to large amounts of diffuse scattering, we only measure reliable results for maximized SH signals at the C resonance. Thus, the flake samples give access to highly resolved SH spectra, whereas absolute values may be extracted from the bulk sample, which, in turn, yield very little spectral information.    
By combining the measured SH response from these two sample types, we are able to calibrate the intensity measured from the exfoliated flakes against the bulk crystal measurements, thereby allowing for estimation of absolute parameters. 
\begin{figure}
	\centering
	\includegraphics[width=1\linewidth]{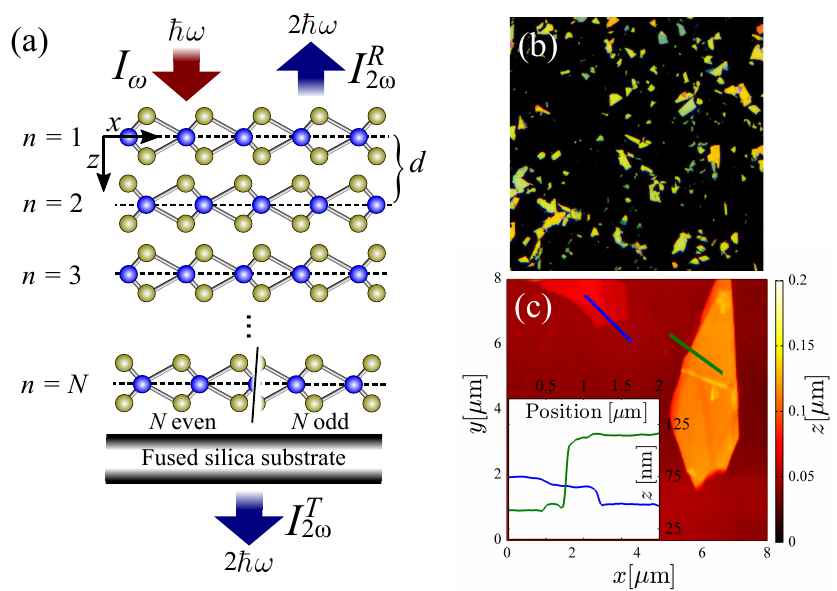}
	\caption{(a) Schematic representation of $N$ 2H-stacked MoS$_2$ layers, each with a sheet SH susceptibility of alternating sign. An incoming fundamental intensity $I_\omega$ induces SH reflected $R^{(2)}$ and transmitted $T^{(2)}$ signals. (b) Microscope image of thick MoS$_2$ flakes exfoliated onto fused silica. The image is 140$\times$140 $\mu$m. (c) Atomic force micrograph of a few selected MoS$_2$ flakes. The inset displays height ($z$) profiles along the lines indicated on the main figure.}
	\label{fig:schematic}
\end{figure}

In Figs. \ref{fig:schematic}(b) and (c), we include microscope images of the produced flakes. Atomic force microscopy and profiler scans confirm most flakes to consist of 40 - 100 monolayers (MLs). Indeed, no measurable photo-luminescence (PL) was observed, regardless of pump wavelength. This suggests very little ML coverage, since only ML domains posses the direct gap necessary for efficient PL\cite{Mak2010,Eda2011,Splendiani2011}. 

At first glance, dipole selection rules prevent generation of even-ordered non-linearities in  MoS$_2$ due to inversion symmetry, which is only broken in finite samples consisting of an odd number of 2H-stacked MoS$_2$ layers\cite{Heinz2013,Kumar2013,Malard2013}. On the other hand, thick flakes, such as those investigated here, are not expected to display properties markedly different from bulk samples.
From a simplistic point of view, the layer-selective vanishing SH response may be understood from the fact that adjacent layers are mirror images of each other in the $yz$-plane, as depicted in Fig. \ref{fig:schematic}(a). Hence, in the limit of weak inter-layer coupling, adjacent layers contribute anti-symmetrically to the SH response function $\chi^{(2)}$.
 While this means that the net polarization of two adjacent MoS$_2$ layers cancels in the dipole approximation, the spatial variation of the driving field makes this cancellation only approximate. 
 Indeed, previous measurements on bulk MoS$_2$ crystals confirm this, with non-vanishing (although very weak) SH signals reported\cite{Wagoner1998,Heinz2013}.
 %
 %
%
%
%
It is this effect that allows us to perform SH spectroscopy on many-layered MoS$_2$ samples. We here analyse the measured response in terms of an effective, spatially dependent SH susceptibility $\chi^{(2)}(z)$, constructed from the effective sheet susceptibilities $\pm \chi_{S,\textrm{eff}}^{(2)}$ of the individual  MLs, viz  
\begin{align}
	\chi^{(2)}(z,\omega) = -\chi_{S,\textrm{eff}}^{(2)}(\omega) \sum_{n=1}^{N}(-1)^{n} \delta(z-nd+d).
	\label{eq:shgresponse}
\end{align}
Here, $d\approx6.2$ Å is the centre-to-centre interlayer distance depicted in Fig. \ref{fig:schematic}(a).

%

%
%

The SH experiments are performed using a Nd:YAG pumped optical parametric oscillator, yielding 5 ns pulses at a repetition rate of 10 Hz and average pulse energies of approximately 1 mJ at the sample. 
Measurements on flake and bulk samples are recorded in transmission and reflection configuration, respectively, using a small optical angle of incidence approximating 4$^\circ$.
Furthermore, the spectral distribution of the emitted radiation was analysed with great care to rule out contributions from competing non-linear processes, such as two-photon PL.    
All results are reported relative to the SH intensity reflected from an $\alpha$-quartz (Qrz) wedge, related to the pump intensity $I_\omega$ by the expression 
$I_{2\omega}^{\mathrm{Qrz}} = |\rho^{(2)}_{\mathrm{Qrz}}|^2 |\chi_{\mathrm{Qrz}}^{(2)}|^2 I_\omega^2$.
 The bulk SH susceptibility of quartz has been measured to\cite{Hagimoto1995} $\chi_{\mathrm{Qrz}}^{(2)}\approx 0.3$ pm/V, while the SH reflection coefficient $\rho^{(2)}_{\mathrm{Qrz}}$ has been derived before\cite{Bloembergen1962}.

We also note that the applied spot size is on the order of 1 mm$^2$, hence, encompassing multiple crystal domains of random orientation. For this reason, we are not able to observe rotational anisotropy, which is routinely seen when investigating few-layered TMDs by SH microscopy\cite{Heinz2013,Malard2013,Kumar2013}.
Further, the various domains are mapped onto separate areas of the PMT detector, which means their respective contributions add incoherently, ruling out interference effects from adjacent crystal grains. 
\begin{figure}
	\centering
	\includegraphics[width=1\linewidth]{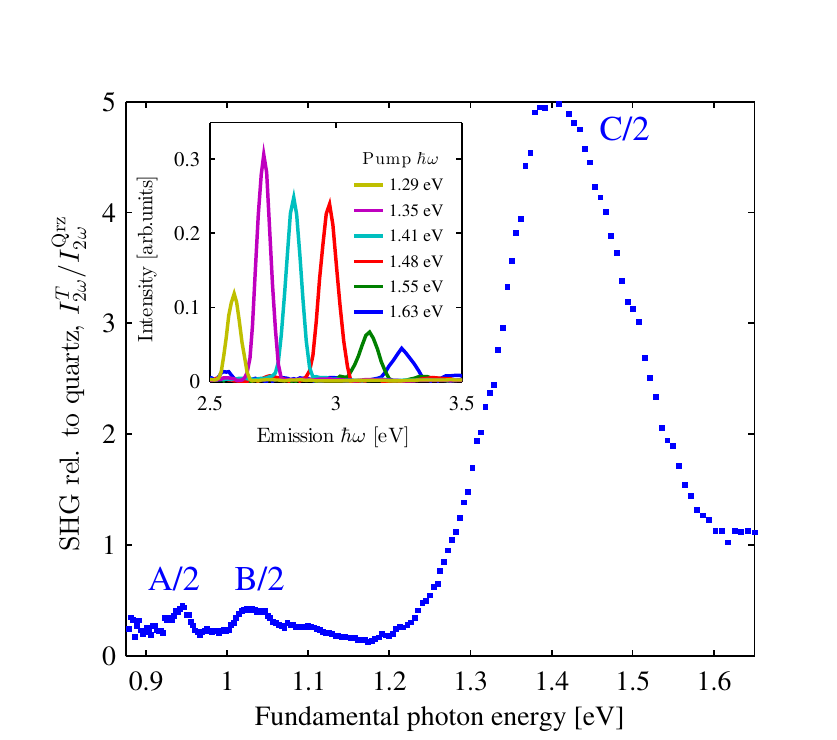}
	\caption{SH signal relative to a quartz reference for thick MoS$_2$ flakes deposited on fused silica. The inset displays the emission spectra for a selection of pump photon energies. No two-photon PL is observed at photon energies corresponding to half the MoS$_2$ fundamental exciton energy.}
	\label{fig:shg}
\end{figure}

In Fig. \ref{fig:shg}, we present the measured 
SH transmission intensity spectrum $I^{T}_{2\omega}$ for a sample with MoS$_2$ flakes.
%
We observe clear maxima in the SH spectrum at fundamental photon energies corresponding to half the energies of the linear absorption features denoted A, B, and C in Ref. \onlinecite{Li2014}. 
 Similar maxima appear in our simulation\cite{Trolle2014} of the excitonic SH response of ML MoS$_2$, where the corresponding peaks in the SH spectrum were dubbed A/2, B/2, and C/2, as also indicated on Fig. \ref{fig:shg}.  
Such ML simulations do not include effects of interlayer coupling, which are obviously present in the samples investigated here. 
	To consider effects of multiple layers, we have extended our exciton model\cite{Trolle2014} to  study  SH generation in trilayer (TL) MoS$_2$, which represents a minimal MoS$_2$ multilayer flake without inversion symmetry. We expand exciton states in a basis of singly excited Slater determinants, which are, in turn, constructed from tight-binding orbitals derived from the parameters of Ref. \onlinecite{Zahid2013}.
	For MLs, the low-energy spectrum is dominated by transitions from the spin-orbit split top valence bands to the degenerate conduction bands near the K-points of the Brillouin zone, giving rise the A and B excitons, as mentioned. In bilayer MoS$_2$, a very similar picture with two split, doubly degenerate top valence bands and a degenerate conduction band minimum near the K-points is observed. Here, however, the splitting is not a spin-orbit effect, but due to crystal field splitting arising from  interlayer coupling. In TL MoS$_2$, we observe a combination of spin-orbit and crystal field splitting, giving rise to two groups of triply near-degenerate valence bands, which preserve the low-energy A/B exciton structure. When adding more layers, transitions in the energy range corresponding to the C peak become dominated by additional bands, which are not degenerate, resulting in a broadened C-feature. See the supplementary material\cite{suppl} for more information.
	We apply a relatively dense $60\times60$ $k$-grid, including 24 valence and 24 conduction bands. Hence, this problem is computationally difficult, and certainly beyond the scope of direct diagonalization. However, by carefully block diagonalizing the BSE based on out-of-plane symmetry, and by using our Lanczos/Haydock approach for calculation of SH response functions\cite{Trolle2014}, we are able to generate converged optical spectra. 
	Furthermore, as described in Ref. \onlinecite{Qiu2013}, linear spectral features are increasingly broadened with increasing photon energy due to electron-phonon scattering. Here, we approximate this effect by implementing a frequency-dependent phenomenological broadening function, which we take to increase linearly with SH photon energies larger than the fundamental exciton energy\cite{suppl}. 
	Note, the simulated TL sheet susceptibility $\chi_S^{(2)}$ is calculated in the dipole approximation, hence, it represents the sheet susceptibility averaged over the three-layer structure while including interlayer coupling. Hence, this quantity is fundamentally different from $\chi_{S,\textrm{eff}}^{(2)}$, which is an effective ML property valid for electronically decoupled layers. 
	
\begin{figure}
		\centering
		\includegraphics[width=1\linewidth]{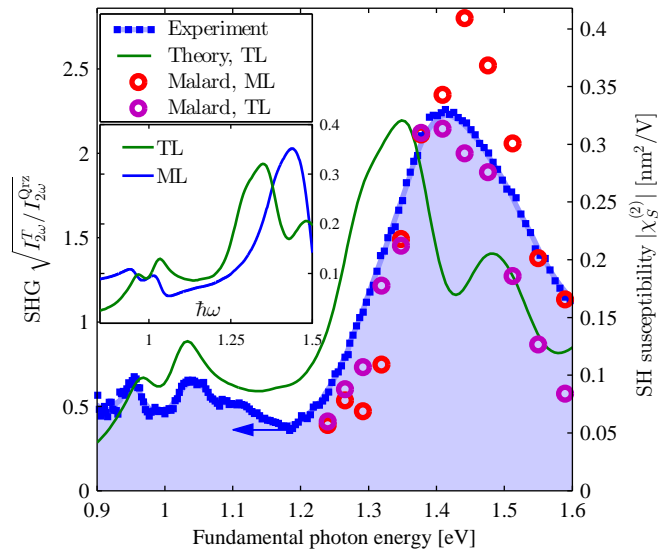}
		\caption{Modulus of the SH response function compared to the square root of the measured SH transmission spectrum. We include also the experimental $|\chi^{(2)}_S|$ spectra for MLs and TLs published by Malard and co-workers\cite{Malard2013}. Note, we scale these spectra by a factor 5 to fit them on the same figure. In the inset we compare our new TL calculation with our previously reported ML spectrum\cite{Trolle2014}, however, now using frequency dependent broadening.}
		\label{fig:shg2}
	\end{figure}

	%
	In Fig. \ref{fig:shg2}, we include the calculated SH response function for TL MoS$_2$, and compare it with the square root of the observed SH signal. 
	%
	The latter quantity is related to the SH response function via $I_{2\omega}^{T} = f^2|\tau_N^{(2)}|^2 |\chi^{(2)}_{S}|^2 I_\omega^2$, where $f$ is the surface coverage and $\tau_N^{(2)}$ is a SH transmission coefficient for an $N$ layer flake\cite{suppl}. As already discussed, both $f$ and $N$ are unknown quantities for the flake samples. However, by assuming $\tau_N^{(2)}$ to depend only weakly on frequency, the square root of the observed SH signal may be compared directly with the SH response function $|\chi_{S}^{(2)}|$. 
	 Indeed, we observe the important A/2, B/2 and C/2 features both in experiments and theory, with similar relative intensities. 
	The theoretical observation of a C/2 splitting absent in the experiments may be attributed to the simplistic implementation of scattering effects. In fact, the linear case is often subject to similar deficiencies\cite{Trolle2014,Qiu2013}. In the inset of Fig. \ref{fig:shg2}, we compare ML and TL results. Although similar trends are found, we clearly observe an effect of including multiple layers, which is more pronounced than for the linear case\cite{suppl}.  
	For the TLs, we observe a slight enhancement of the A/2 and B/2 features relative to the ML case. 
	Furthermore, in the studies using  SH spectroscopy Malard and co-workers\cite{Malard2013} have demonstrated how the C/2 peak in ML samples is slightly red-shifted and broadened in TL samples (we include the spectra of Malard \textit{et al.}\cite{Malard2013} in Fig. \ref{fig:shg2} for comparison). The theoretical results shown in the inset reproduce this trend. 
	
We note that the observed SH intensity, displayed in Fig. \ref{fig:shg}, is roughly five times larger than the quartz reference signal, even though the sample was covered by roughly 10\% MoS$_2$ flakes. Hence, this hints at the extraordinarily large\cite{Malard2013,Kumar2013,Heinz2013} SH response of few-layered MoS$_2$. 
Unfortunately, we are not able to extract absolute values of the SH susceptibilities from measurements on flake samples due to uncertain surface coverages and flake thicknesses.
%
 However, our measurements on a  $\sim$ 5 mm thick bulk MoS$_2$ sample are not affected by thickness due to absorption.
  For fundamental photon energies near 1.4 eV, i.e. at the C/2 resonance, we measure a SH intensity reflected the bulk sample approximately 60 times larger than the SH quartz reference signal. 
To extract an effective sheet susceptibility from this value, we note that the intensity of the reflected SH signal $I_{2\omega}^{R}$ may be related to the fundamental intensity $I_\omega$ and $\chi_{S,\textrm{eff}}^{(2)}$ by the relation $I_{2\omega}^{R} = |\rho^{(2)}_N|^2 |\chi_{S,\textrm{eff}}^{(2)}|^2I_\omega^2$, where $\rho^{(2)}_N$ is a SH reflection coefficient valid for an $N$-layer structure.
 The reflection coefficient $\rho_N^{(2)}$, suitable for an inhomogeneous non-linear response given by Eq. \ref{eq:shgresponse}, may be found using a relatively compact expression\cite{suppl}, which
in the bulk $(N\to\infty)$ and ML $(N=1)$ limits yield 
\begin{align}
	\rho^{(2)}_{\infty}	 &\approx    \frac{4ik_0}{[ 1 +n_{2\omega} ][1+ n_\omega]^2}, \label{eq:Rinf} \\
	\rho^{(2)}_{1} 
	&\approx   \frac{8ik_0}{[1+\tilde n_{2\omega}][1 +\tilde n_\omega]^2 }. \label{eq:R1}
\end{align}
%
Here, $k_0=\omega/c$ while the optical phase shift upon propagating through a single ML is taken to be negligible compared to the shift across all $N$ layers. Furthermore $\tilde n_\omega$ and $n_\omega$ are, respectively, the refractive indices of fused silica and MoS$_2$\cite{Malitson1965,Li2014}. We note that the ML limit Eq. \ref{eq:R1} is identical to the response derived starting from a non-linear surface $\delta$-polarization\cite{Sipe1987,Shen1989}, and has been applied before for the study of few-layered TMDs\cite{Malard2013,Heinz2013}. Also, the bulk to ML intensity ratio at $\hbar\omega = 1.4$ eV may be evaluated to  $|\rho^{(2)}_{\infty}/\rho^{(2)}_{1}|^2  \approx 0.15$\% , agreeing favourably with previously reported results\cite{Heinz2013}. 
	It is also noted that the bulk expression Eq. \ref{eq:Rinf} is very similar to the corresponding result for SH reflection from an infinite medium with homogeneous non-linear response\cite{Bloembergen1962}, with the main difference being the $k_0$ proportionality, reflecting the fact that $k_0 \chi^{(2)}_{S,\textrm{eff}}$ acts as an effective bulk susceptibility. 
Since we measure the ratio $I_{2\omega}^{R}/I_{2\omega}^{\textrm{Qrz}}=|\chi_{S,\mathrm{eff}}^{(2)} \rho^{(2)}_\infty|^2/|\chi_{\mathrm{Qrz}}^{(2)} \rho^{(2)}_{\mathrm{Qrz}}|^2$, and since we have expressions for both $\rho^{(2)}_\infty$ and $\rho^{(2)}_{\textrm{Qrz}}$ in addition to $\chi_{\textrm{Qrz}}^{(2)}$, we 
	may estimate an effective ML sheet susceptibility defined previously in Eq. \ref{eq:shgresponse} to $|\chi_{S,\mathrm{eff}}^{(2)}| \approx 1.5$ nm$^2$/V.
	
	Based on our measurements, we find the effective sheet SH susceptibility $|\chi_S^{(2)}|$ to be somewhat larger than the $\sim 0.1$ nm$^2$/V reported for MoS$_2$ MLs on quartz substrate in Refs. \onlinecite{Malard2013} and \onlinecite{Heinz2013}. On the other hand, in Ref. \onlinecite{Kumar2013}  susceptibilities as large as 60 nm$^2$/V are reported. In addition to the experimental values mentioned, on-resonance moduli of $\chi_S^{(2)}$ derived from theory are typically in the $\sim$ nm$^2$/V range\cite{Trolle2014,Gruning2014,Guo2014}. Hence, the absolute value of the SH response function of MoS$_2$ is still a topic of considerable debate. We note that results generated using advanced microscopy techniques for resolution of individual TMD flakes with well-defined thickness, and by applying well-documented reference measurements for extraction of absolute values, such as Refs. \onlinecite{Heinz2013,Malard2013}, should be regarded with a high degree of validity. 
	
	The bulk crystal applied in our work is composed of multiple grains, with different orientations. On the boundaries of such grains, the odd/even layer $\chi^{(2)}_S$ anti-symmetry, implying dipole cancellation for perfectly stacked samples, does not lead to complete cancellation due to misalignment. Hence, these domains might give rise to an additional SH signal, perhaps partly explaining the large response measured here.

In conclusion, we have extended the experimentally available SH spectrum of multi-layered MoS$_2$ flakes to include the broad fundamental photon energy range between 0.85 and 1.7 eV. Moreover, we have performed an exciton simulation of the nonlinear optical properties of TL MoS$_2$, and find the key features of the simulated spectrum to be in good agreement with the experimentally observed spectrum. Hence, in this work, we have identified experimentally the signatures of the A, B and C excitons, in addition to the relative intensities of the resulting peaks. Hence, these results extend the spectral region in which experimental data are now available for comparison with theoretically derived SH spectra, allowing for future comparisons with more advanced models.   

\section*{Acknowledgements}
The authors gratefully acknowledge the financial support from the Villum Foundation Center of Excellence QUSCOPE. The authors also express their gratitude to the Centre of Scientific Computing Aarhus (Aarhus University) for access to the Grendel computer cluster. Furthermore, the authors would like to thank Bjarke Jessen at the Technical University of Denmark for supplying the investigated MoS$_2$ flakes. Additionally, we also gratefully acknowledge the authors of Ref. \onlinecite{Malard2013} for supplying their experimental results, which are reproduced in Fig. \ref{fig:shg2} for comparison.

\bibliography{bib}

\end{document}